\documentstyle[12pt,twoside]{article}
\def\greaterthansquiggle{\raise.3ex\hbox{$>$\kern-.75em\lower1ex\hbox{$\sim$}}}
\def\lessthansquiggle{\raise.3ex\hbox{$<$\kern-.75em\lower1ex\hbox{$\sim$}}}
\newcommand{\beq}{\begin{equation}}
\newcommand{\eeq}{\end{equation}}
\newcommand{\beqa}{\begin{eqnarray}}
\newcommand{\eeqa}{\end{eqnarray}}
\newcommand{\beqan}{\begin{eqnarray*}}
\newcommand{\eeqan}{\end{eqnarray*}}
\newcommand{\ba}{\begin{array}}
\newcommand{\ea}{\end{array}}
\newcommand{\no}{\nonumber}

\newcommand{\ol}{\overline}

\newcommand{\wt}{\widetilde}

\newcommand{\C}{{\cal C}}

\newcommand{\R}{{\cal R}}

\def\nz{\ifmmode {I\hskip -3pt N} \else {\hbox {$I\hskip -3pt N$}}\fi}
\def\zz{\ifmmode {Z\hskip -4.8pt Z} \else
       {\hbox {$Z\hskip -4.8pt Z$}}\fi}
\def\qz{\ifmmode {Q\hskip -5.0pt\vrule height6.0pt depth 0pt
       \hskip 6pt} \else {\hbox
       {$Q\hskip -5.0pt\vrule height6.0pt depth 0pt\hskip 6pt$}}\fi}
\def\rz{\ifmmode {I\hskip -3pt R} \else {\hbox {$I\hskip -3pt R$}}\fi}
\def\cz{\ifmmode {C\hskip -4.8pt\vrule height5.8pt\hskip 6.3pt} \else
       {\hbox {$C\hskip -4.8pt\vrule height5.8pt\hskip 6.3pt$}}\fi}

\def\au{{\setbox0=\hbox{\lower1.36775ex%
\hbox{''}\kern-.05em}\dp0=.36775ex\hskip0pt\box0}}
\def\ao{{}\kern-.10em\hbox{``}}

\normalsize
\begin{document}
\bibliographystyle{plain}
\begin{titlepage}
\begin{flushright}
UWThPh-1999-53
\end{flushright}
\vspace{2.5cm}
\begin{center}
{\Large \bf Multipole Moments of Static Spacetimes}\\[50pt]
R. Beig*   \\
Institut f\"ur Theoretische Physik \\ Universit\"at Wien\\
Boltzmanngasse 5, A-1090 Wien 
\vfill
\end{center}

Talk given at Arthur L. Besse, Table Ronde de G\'eometrie 
Pseudo-Riemannienne Globale, Institut \'Elie Cartan, Universit\'e Henri
Poincar\'e -- Nancy 1
\vfill
\noindent
*) Supported in part by Fonds zur F\"orderung der wissenschaftlichen Forschung,
Project No. P12626-PHY.
\end{titlepage}

In this talk I describe work, mostly done in collaboration with W. Simon
some 20 years ago [1,2], on multipole moments of static spacetimes. My
purpose is to make this work, which lies at the interface of classical
potential theory, conformal geometry and general relativity, known to
mathematicians and to perhaps motivate them to have a look at the open
problems which still remain.

Our basic object will be an asymptotically flat, smooth or analytic
Riemannian 3-manifold $(\wt M,\wt g_{ij})$ where $\wt M$ is diffeomorphic
to ${\bf R}^3 \setminus {\bf B}_{R_0}$ with ${\bf B}_{R_0}$ the Euclidean
ball of radius $R_0$. In the chart given by this diffeomorphism the metric
$\wt g_{ij}$ is required to satisfy the fall-off conditions
\beqa
\wt g_{ij} - \delta_{ij} &=& O(1/r), \qquad r^2 = \delta_{ij} x^i x^j \no \\
\partial \wt g_{ij} &=& O(1/r^2) \\
\partial^2 \wt g_{ij} &=& O(1/r^3). \no
\eeqa
It is often convenient and sometimes vital to have a stronger notion of
asymptotic flatness which is defined via ``conformal compactification''
(CC) and goes as follows: There should exist a positive function $\Omega$
on $\wt M$, such that the metric $g_{ij} = \Omega^2 \wt g_{ij}$ extends
to a smooth, or perhaps analytic metric on $M = \wt M \cup \{r = \infty\} =
\wt M \cup \{\Lambda\}$, where
\beq
\Omega|_\Lambda = 0, \qquad
\left.\Omega_i \right|_\Lambda = 0, \qquad
\left. (\Omega_{ij} - 2 g_{ij})\right|_\Lambda = 0,
\eeq
with $\Omega_{i \ldots}$ denoting covariant derivatives. There is a gauge
freedom associated with this definition. Namely, if $\Omega$ gives a CC,
then $\ol{\Omega} = \omega \Omega$, with $\omega|_\Lambda = 1$, gives
another one. Note that, with the above notion of asymptotic flatness,
there exists a CC with a rescaled (``unphysical'') metric $g_{ij}$ being
$C^0$ but not $C^1$ in general. Namely, pick $\Omega = 1/r^2$ and
\beq
x'{}^i = x^i/r^2 \qquad \mbox{(``Kelvin inversion'')}
\eeq
as coordinates on $M$. Then
\beq
g'_{ij} = (\delta^k{}_i - n'{}^k n'_i)(\delta^\ell{}_j - n'{}^\ell
n'_j) \wt g_{k\ell}
\eeq
with $n'{}^i = x'{}^i/r'{}^2$, $n'_i = \delta_{ij} n'{}^j$. Hence
\beqa
g'_{ij} &=& \delta_{ij} + (\delta^k{}_i - n'{}^k n'_i)(\delta^\ell{}_j
- n'{}^\ell n'_j)(\wt g_{k\ell} - \delta_{k\ell}) \no \\
&=& \delta_{ij} + O(r').
\eeqa
Clearly, if $\wt g_{ij}$ is flat, $g_{ij}$ is also flat, so there is a
smooth, even analytic, CC in this case. In some applications where a
smooth CC exists, the conformal factor $\Omega$ can not be chosen to be
smooth. As an important example take the spatial Schwarzschild metric.
This is given by
\beq
\wt g_{ij} dx^i dx^j = \left( 1 + \frac{m}{2r}\right)^4 \delta_{ij}
dx^i dx^j \qquad (m > 0).
\eeq
Then, taking $x'{}^i = x^i/r^2$ and $\Omega = r'{}^2(1 + mr'/2)^2$, we
have that $g_{ij} = \delta_{ij}$, but $\Omega$ is only $C^2$. Clearly,
if we had started from $\delta_{ij}$ rather than $\wt g_{ij}$ as the
basic metric this problem would not have arisen. Fortunately, 
if $\wt g_{ij}$ is the spatial metric of an asymptotically flat static
spacetime satisfying the Einstein vacuum equations, as is the
Schwarzschild metric, there is a general
way of writing $\Omega = f \Omega'$ so that $\Omega'$ is smooth. An
asymptotically flat vacuum spacetime, for the present purposes, is the
manifold $({\bf R} \times \wt M,ds^2)$ with $ds^2$ given by
\beq
ds^2 = - e^{2 \wt U} dt^2 + \wt g_{ij} dx^i dx^j,
\eeq
with $\wt U$ a function on $\wt M$ satisfying
\beq
\wt U = O(1/r), \qquad \partial \wt U = O(1/r^2), \qquad
\partial^2 \wt U = O(1/r^3).
\eeq
We now rewrite $\wt g_{ij}$ as
\beq
\wt g_{ij} = e^{-2 \wt U} \wt \gamma_{ij}.
\eeq
In the case of the Schwarzschild metric we have that
\beq
e^{2 \wt U} = \left(\frac{1 - m/2r}{1 + m/2r}\right)^4, \qquad
\wt \gamma_{ij} = \left( 1 - \frac{m^2}{4r^2} \right)^2 \delta_{ij},
\qquad r > m/2.
\eeq
(Thus $R_0$ has to be taken $\geq m/2$.) It easily follows that
$\wt \gamma_{ij}$ has a smooth CC with smooth conformal factor. The
Schwarzschild solution is a solution of the Einstein vacuum equations on
${\bf R} \times \wt M$. These equations are equivalent to
\beq
\Delta_{\wt \gamma} \wt U = 0,
\eeq
\beq
\R_{ij} [\wt \gamma] = 2 \wt U_i \wt U_j,
\eeq
where $\Delta_{\wt \gamma}$ is the Laplace Beltrami operator on
$(\wt M, \wt \gamma_{ij})$ and $\R_{ij}[\wt \gamma]$ the Ricci tensor of
$\wt \gamma_{ij}$. The equations (11,12) should be viewed as the
``source-free static Einstein equations on an exterior domain''. In
particular, $(\wt U, \wt \gamma_{ij})$ could have smooth extensions to
${\bf R}^3$ so that $ds^2$ satisfies the static Einstein equations on
${\bf R} \times {\bf R}^3$ with an energy momentum tensor corresponding
to some ``reasonable'' matter model. If, for example, this matter model
corresponds to a perfect fluid with density $\rho$, pressure $p$ and an
equation of state of the form $p = p(\rho)$ satisfying a certain
differential inequality, it is known [3,4] that the spacetime is
spherically symmetric (whence Schwarzschild in ${\bf R} \times \wt M$),
and thus there is just a one-parameter family of solutions, parametrized,
say, by the central density. In the trivial case that the spacetime is
vacuum throughout ${\bf R} \times {\bf R}^3$ it is known [5] that
$ds^2$ is the Minkowski metric. (If the matter model describes an
elastic solid such as the earth, the solutions will in general be neither
Schwarzschild nor flat space, of course.) The point of restricting
ourselves to an exterior domain is that we do not care about what sort
of matter may be ``inside''. Then there is a larger class of solutions.
How large, in fact, this class is and how it can be conveniently
characterized,
is precisely the question we are asking. On physical grounds one expects
that there should be the same freedom as in the Newtonian theory, to
which we now turn. The Equ.'s (11,12) are now replaced by 
\beqa
\Delta_{\wt \gamma} \wt U &=& 0, \qquad \wt U = O(1/r) \\
\R_{ij}[\wt \gamma] &=& 0.
\eeqa
Thus, since $\dim \wt M = 3$, the Riemann tensor of $\wt \gamma_{ij}$ has
to vanish. By the simple connectedness of $\wt M$ it follows that
$\wt \gamma_{ij} = \delta_{ij}$ in suitable coordinates, and thus there
remains
\beq
\Delta_\delta  \wt U = 0,
\eeq
that is to say the Laplace equation with respect to the Euclidean metric.
It is convenient to replace (15) by
\beq
\Delta_\delta  \wt U = - 4 \pi \rho,
\eeq
where $\rho \in C^\infty_0({\bf R}^3)$, whence $\wt U$ has to be of the form
\beq
\wt U(x) = \int_{{\bf R}^3} \frac{\rho(x')}{|x-x'|} dx'.
\eeq
It follows [6] that $\wt U$ has an expansion of the form
\beq
\wt U = \frac{M}{r} + \frac{M_ix^i}{r^3} + \frac{M_{ij}x^ix^j}{2r^5} +
\ldots + \frac{M_{i_1\ldots i_k} x^{i_1} \ldots x^{i_k}}{k!   r^{2k+1}}
+ \ldots
\eeq
where the constants $M_{i_1 \ldots i_k}$ are symmetric and tracefree,
which converges uniformly in $\wt M$ (see [6]). As a warm-up for general
relativity we outline an independent argument for this classical result,
as follows: One first shows that
\beq
\wt U = \sum_{\ell = 0}^k \frac{M_{i_1 \ldots i_\ell} x^{i_1} \ldots
x^{i_\ell}}{\ell!  r^{2\ell  + 1}} + O(1/r^{k+1})
\eeq
for some sufficiently large $k$, say $k = 2$, and that this relation can be
differentiated at least twice (i.e. any partial derivative of the 
left-hand side is $O(1/r^{k+2})$, a.s.o.). Then pick the standard CC.
Now define an unphysical potential $U$ by $U=\Omega^{-1/2} \wt U$. This
has the form
\beq
U = M + M_i x'{}^i + \frac{1}{2} M_{ij} x'{}^i x'{}^j + O(r'{}^3).
\eeq
Thus $U$ is $C^{2,\alpha}$ ($0 < \alpha \leq 1$) in $B_{1/R} = M$.
Next observe that $U$ again satisfies the Laplace equation. This is seen
by first recalling the conformal behaviour of the conformal Laplacian, i.e.
\beq
\left(\Delta_\gamma - \frac{\R}{8} \right) U = \Omega^{-5/2} \left(
\Delta_{\wt \gamma} - \frac{\wt \R}{8} \right) \wt U, \qquad
U=\Omega^{-1/2} \wt U
\eeq
and specializing to the case where both $\wt \gamma_{ij}$ and $\gamma_{ij}$
are flat. We now have that $U$ is a $C^{2,\alpha}$-solution of
\beq
\Delta_\delta  U = 0 \qquad \mbox{in } M.
\eeq
But it is well known that solutions to the Laplace equation have to be
analytic [7]. Thus $U$ has a convergent Taylor expansion at the origin, 
which ends the proof.

Since analyticity, by [8], is a general property of solutions to elliptic
systems with analytic coefficients, one can try to use a similar line of
thought for the relativistic case. The main obstacle is that the equations
(11,12), far from being invariant under the conformal rescaling
\beq
U = \Omega^{-1/2} \wt U, \qquad \gamma_{ij} = \Omega^2 \wt \gamma_{ij},
\eeq
become formally singular at the point $\Lambda$ where $\Omega$ vanishes.
Before showing how this obstacle can be overcome we have to perform the
analogue of the first step in the Newtonian case, namely the analysis of
the first few terms of the expansion of the quantities $(\wt U,
\wt \gamma_{ij})$ in powers of $1/r$. This has been done, up to arbitrary
orders in $1/r$, in the sequence of works [9,10,1]. One could obtain 
sufficiently detailed information about the general term in order to show:
For every $k$ there exist coordinates $x^i$ so that, after the standard CC,
$\gamma_{ij}$ is smooth. For example, for $k = 3$, one finds
\beq
\wt \gamma_{ij} = \delta_{ij} - \frac{M^2}{r^4} (\delta_{ij}r^2 - x_ix_j)
- \frac{2M M_{(i}x_{j)}}{r^4} + \frac{2MM_k x^k}{r^6}
(-\delta_{ij}r^2 + 2x_ix_j) + O(1/r^4),
\eeq
\beq
\wt U = \frac{M}{r} + \frac{M_ix^i}{r^3} + \frac{1}{2}
\frac{M_{ij}x^ix^j}{r^5} + \frac{M^3}{r^3} + O(1/r^4)
\eeq
for some constants $M$, $M_i$, $M_{ij}$ ($M_{ij}$ symmetric and tracefree)
and where $x_i = \delta_{ij} x^j$. These equations show in particular that, at 
least at this order, the relativistic solution has no more free parameters
than the Newtonian one, and this remains true for arbitrary high orders.

Very loosely speaking, whereas there are harmonic functions solving Equ. (11) 
with the appropriate boundary conditions which are parametrized by multipole 
moments, the metric $\wt \gamma$ is essentially determined by the r.h. side of 
Equ. (12). This, in turn, is related to the fact that $\R_{ij}$ determines
the curvature tensor in 3 dimensions and that zero curvature implies a flat
metric. Within the iterative scheme by which the system (11,12) is solved
this means that, for each order in $1/r$, ``harmonic'' contributions to
$\wt \gamma_{ij}$ have to be pure gauge.

Armed with this information one can now study a suitable ``unphysical'' 
version of Equ.'s (11,12). It is not clear how to use the standard CC
for this purpose. The basic observation is that the expansion of $\wt U$,
as in (25), also says that $\wt U^2$ is arbitrarily smooth in Kelvin
inverted coordinates and that, when $M \neq 0$, $(\Omega,\gamma_{ij})$
satisfies conditions (2) with $\Omega = \wt U^2/M^2$. It is instructive
to first see the effect of this CC in the Newtonian case. In that case we
have that $\wt \R = 0$ so that $\wt \Delta \wt U = 0$ and $U \equiv M$.
Now Equ. (21) implies
\beq
\R = 0.
\eeq
Conversely we find from $(\wt \Delta - \wt \R/8) \cdot \mbox{const} = 0$
that, outside $\Lambda$, $\Delta(\Omega^{-1}) = 0$. Equivalently
\beq
\Omega \Delta \Omega = \frac{3}{2} \Omega_i \Omega^i.
\eeq
Of course, the unphysical metric $\gamma_{ij}$ is now not flat any longer
in general. It satisfies
\beq
- \Omega \R_{ij} = \left(\Omega_{ij} - \frac{1}{3} g_{ij} \Delta 
\Omega\right).
\eeq
The relation (27), as it stands, does not give rise to a regular elliptic
equation for $\Omega$ since
\beq
\sigma := \frac{3}{2} \frac{\Omega_i \Omega^i}{\Omega}
\eeq
is formally singular at $\Lambda$. Using (2.8) we find that
\beq
\sigma_i = - 3 \R_{ij} \Omega^j
\eeq
which further implies
\beq
\Delta \sigma = 3 \Omega \R_{ij} \R^{ij}.
\eeq
Now (31), together with $\Delta \Omega = \sigma$, {\em does\/} form a
regular elliptic system. Using the theorem of Morrey [8], it would 
now follow that $(\Omega,\sigma)$ are analytic provided $\gamma_{ij}$ is.
This, in turn, follows from the following differential geometric
\paragraph{Lemma:} A conformally flat metric with zero scalar curvature is 
analytic.
\paragraph{Proof:} By $\R = 0$ and conformal flatness there holds
\beq
D_k \R_{ij} = D_i \R_{kj}.
\eeq
Taking $D^k$ of Equ. (32), commuting derivatives and using
\beq
\R_{kji\ell} = 2 \R_{i[k} g_{j]\ell} - 2 \R_{\ell[k} g_{j]i}
\eeq
and the Bianchi identity we obtain
\beq
\Delta \R_{ij} = 3 \R_i{}^k \R_{jk} + g_{ij} \R_{k\ell} \R^{k\ell}.
\eeq
Now recalling that the Ricci tensor of a metric $g_{ij}$, when written
in harmonic coordinates, gives an elliptic operator for $\gamma_{ij}$,
we see that
\beq
\R_{ij} = \sigma_{ij},
\eeq
together with (34) yields an elliptic system for $(\gamma_{ij},
\sigma_{ij})$ with analytic coefficients. Thus, by [8], the Lemma is
proved.

In the case at hand we have that $\R$ vanishes and, furthermore, that
$\gamma_{ij}$ is conformally flat. In fact, from (29) we can directly
obtain
\beq
\R_{i[j;k]} = 0.
\eeq
Combining the analyticity of $\gamma_{ij}$ with the system (30,31) we
obtain analyticity of the pair $(\Omega,\gamma_{ij})$.

The relativistic situation, given by Equ.'s (13,14), is similar, but 
more complicated. Instead of
$\Omega = \wt U^2/M^2$ it is more convenient to take
\beq
\Omega = \frac{4}{M^2} \sinh^2 \frac{\wt U}{2}
\eeq
as conformal factor. As a field variable it is useful to take, instead of
$\Omega$, the quantity $\rho$, defined by
$$
\rho = \tanh^2 \frac{\wt U}{2}. \eqno(37')
$$
Then, again, it turns out that $\R = 0$ and
\beq
\Delta \rho = \sigma := \frac{3}{2} \frac{\rho_i \rho^i}{\rho},
\eeq
whereas (28) gets replaced by
\beq
- \rho \left(1 - \rho \right) \R_{ij} =
\rho_{ij} - \frac{1}{3} \gamma_{ij} \Delta \rho.
\eeq
Clearly the metric $\gamma_{ij}$ can not any longer be conformally flat.
In fact, taking the ``curl'' of (39), Equ. (36) of the Newtonian theory
gets replaced by
\beq
 \left(1 - \rho\right) \R_{i[j;k]} =
2\R_{i[j} \rho_{k]} + \gamma_{i[j} \R_{k]\ell} \rho^\ell.
\eeq
As a side remark we add,that, from the above equation it is not hard to
show that $\gamma_{ij}$ is
conformally flat iff $(\Omega,\gamma_{ij})$ comes from the Schwarzschild
solution.

For $\sigma$ we obtain, instead of (31), the relation
\beq
\Delta\sigma = 3 \rho \left(1 - \rho \right)^2+ 3 
\R_{ij} \rho^i \rho^j.
\eeq
Apropos Equ. (41) we wish to point out that, with the substitution
\beq
\kappa := \frac{\sigma^{1/4}}{(1 - \rho)^{1/2}}
\eeq
and using (40), there results the identity
\beq
\Delta \kappa = \frac{1}{2} \kappa^5 + \frac{9}{16} \kappa^{-7} \R_{i[j;k]}
\R^{i[j;k]},
\eeq
which plays an important role in black hole uniqueness theory (see e.g.
[11]).

In any case, we can now consider the system consisting of (35), (38), (41) 
and the equation of the form
\beq
\Delta \sigma_{ij} = \ldots
\eeq
obtained by multiplying (40) by $(1 - \rho)^{-1}$ (which is non-zero
near $\Lambda$) and taking $D^k$. This
furnishes an elliptic system with analytic coefficients for the variables
$(\rho, \gamma_{ij},\sigma,\sigma_{ij})$ (that-is-to-say if harmonic
coordinates are used in Equ. (35)). We conclude that $(\Omega,\gamma_{ij})$
is analytic.

In the Newtonian case, the multipole moments of the solution were nothing
but the Taylor coefficients at $\Lambda$ of the unphysical potential $U$ in
the standard CC. In particular this implies that these moments determine the 
physical solution uniquely (which however already follows from the expansion
(18)). In the relativistic case, and using the CC above based on
$\Omega = \wt U^2/M^2$, it is not immediately clear whether the moments
determine the solution. In fact, the expansions in (24,25) use a 
specific coordinate condition. In the unphysical picture one would like a 
``gauge invariant'' definition of multipole moments. Luckily, Geroch in [12]
came up with such a definition. It goes as follows: Consider the following
recursively defined set of tensor fields built from $(U,\Omega,\gamma_{ij})$
\beqa
P &=& U \no \\
P_i &=& U_i \no \\
P_{ij} &=& U_{ij} - \frac{1}{3} \gamma_{ij} \Delta U \\
P_{i_1 \ldots i_{s+1}} &=& \C \left[ D_{i_{s+1}} P_{i_1 \ldots i_s} -
\frac{s(2s-1)}{2} \R_{i_1i_2} P_{i_3 \ldots i_{s+1}}\right], \no
\eeqa
where $\C$ denotes the operation of taking the tracefree symmetric part. 
Now define
\beq
M_{i_1 \ldots i_s} := \left. P_{i_1 \ldots i_s}\right|_\Lambda.
\eeq
Clearly, in the Newtonian case and using the standard CC for which
$\R_{ij} = 0$, this definition coincides with the standard one. As for the
gauge dependence, when
\beq
\ol{\Omega} = \omega \Omega, \qquad
\ol{U} = \omega^{-1/2} U, \qquad
\ol{\gamma}_{ij} = \omega^2 \gamma_{ij},
\eeq
with $\left. \omega\right|_\Lambda = 1$ it turns out [13] that
\beq
\ol{M}_{i_1 \ldots i_s} = M_{i_1 \ldots i_s} + \C \sum_{r=0}^{s-1}
\left( \ba{c} s \\ r \ea \right) (2s-1) \ldots (2r+1)(-1)^{r-s}
M_{i_1 \ldots i_s} b_{i_{r+1}} \ldots b_{i_s},
\eeq
where $b_i := \left.\frac{1}{2} \omega_i \right|_\Lambda$. Note that 
only first derivatives of $\omega$ at $\Lambda$ enter the transformation
law (48). Furthermore the dependence of $M_{i_1 \ldots i_s}$ on $b_i$
is exactly the same as the change of the Newtonian moments under a
translation of the physical Euclidean space by the vector $b_i$. We can
now come back to the conformal gauge with $\Omega = \wt U^2/M^2$. Here
it follows from (45) that $U \equiv M$ so that the moments are
equivalent to $M$, together with $\C$ of the derivatives of $\R_{ij}$
at $\Lambda$. But it is not difficult to infer from the above equations
that these data determine $(\Omega,\gamma_{ij})$ uniquely whence
$(\wt U,\wt \gamma_{ij})$ is also determined. We remark, finally, that
the $M_{i_1 \ldots i_s}$-terms in the physical expansion, which we have
written out for $k=2$ in (24,25), coincide with the Geroch moments from
above. 

There are two problems which the above analysis leaves open. The first one
is to give an analyticity proof in the case where $M$ is zero. Although
the ``physical'' expansion of $(\wt U,\wt \gamma_{ij})$ seems to be
completely insensitive to whether $M$ is nonzero or not, the unphysical
situation, because of the choice of conformal factor makes vital use of
the nonvanishing of $M$. The second open issue is that of convergence.
Namely, given a set of moments $M_{i_1 \ldots i_s}$, $s = 0,1,\ldots$:
what is the condition on the behaviour for large $s$ so that a static
vacuum solution having these as multipole moments exists? The Newtonian
situation is clear: the moments have to be such that the Taylor series
having them as coefficients converges. In the relativistic case, the
Taylor coefficients at $\Lambda$ of the quantities $(\Omega,\gamma_{ij})$,
say in Riemannian normal coordinates, depend on these moments in a
nonlinear fashion. In particular it is not even obvious whether a solution
exists for which only finitely many moments are nonzero.


\begin{thebibliography}{99}
\bibitem{1} R. Beig, W. Simon, Commun. Math. Phys. {\bf 78}, 75 (1980)
\bibitem{2} W. Simon, R. Beig, J. Math. Phys. {\bf 24}, 1163 (1983)
\bibitem{3} R. Beig, W. Simon, Commun. Math. Phys. {\bf 144}, 373 (1992)
\bibitem{4} L. Lindblom, A.K.M. Masood-ul-Alam, Commun. Math. Phys.
{\bf 162}, 123 (1994)
\bibitem{5} A. Lichn\'erowicz, Th\'eorie relativistes de la gravitation et
de l'\'electromagnetism, Masson, Paris, 1935
\bibitem{6} O.D. Kellogg, Foundations of Potential Theory, Springer,
Berlin,
1926
\bibitem{7} D. Gilbarg, N. Trudinger, Elliptic Partial Differential
Equations of Second Order, Springer, New York, 1983
\bibitem{8} C.B. Morrey, Am. J. Math. {\bf 80}, 198 (1958)
\bibitem{9} R. Beig, Gen. Rel. Grav. {\bf 12}, 439 (1980)
\bibitem{10} R. Beig, W. Simon, Gen. Rel. Grav. {\bf 12}, 1003 (1982)
\bibitem{11} D.C. Robinson, Gen. Rel. Grav. {\bf 8}, 695 (1977)
\bibitem{12} R. Geroch, J. Math. Phys. {\bf 11}, 2580 (1970)
\bibitem{13} R. Beig, Acta Phys. Austr. {\bf 53}, 249 (1981)
\end{thebibliography}
\end{document}